\documentclass[prl,showpacs,byrevtex,amsmath,twocolumn,amssymb,nofootinbib]{
revtex4}
\usepackage{amsmath,graphicx}
\usepackage{epsfig}
\usepackage{float}

\begin{document}

\title{Capillary Rise in Nanopores: Molecular Dynamics \\
Evidence for the Lucas-Washburn Equation}

\author{D. I. Dimitrov$^1$, A. Milchev$^{1,2}$, and K. Binder$^1$ \\
[\baselineskip]
\it$^{(1)}$ {\it Institut f\"ur Physik, Johannes Gutenberg Universit\"at Mainz, }\\
{\it Staudinger Weg 7, 55099 Mainz, Germany}\\
$^{(2)}$ {\it Institute for Chemical Physics, Bulgarian Academy of Sciences,} \\
{\it 1113 Sofia, Bulgaria} \\
}

\begin{abstract}
When a capillary is inserted into a liquid, the liquid will
rapidly flow into it. This phenomenon, well studied and understood
on the macroscale, is investigated by Molecular Dynamics
simulations for coarse-grained models of nanotubes. Both a simple
Lennard-Jones fluid and a model for a polymer melt are considered.
In both cases after a transient period (of a few nanoseconds) the
meniscus rises according to a $\sqrt{\textrm{time}}$-law. For the 
polymer melt, however, we find that the capillary flow exhibits a 
slip length $\delta$, comparable in size with the nanotube radius 
$R$. We show that a consistent description of the imbibition process
in nanotubes is only possible upon modification of the Lucas-Washburn 
law which takes explicitly into account the slip length $\delta$. 
We also demonstrate that the velocity field of the rising fluid close 
to the interface is not a simple diffusive spreading.
\end{abstract}


\maketitle
\paragraph*{Introduction}
Understanding fluid flow on the nanoscale \cite{1} is crucial for
modern developments of nanotechnology like the ``lab on a chip''
and related nanofluidic devices, as well as for various
applications of porous materials, fluid flow through pores in
biomembranes, etc. A key process is the ability of fluids to
penetrate into fine pores with wettable walls. Filling hollow 
carbon nanotubes or alumina nanopore arrays with chosen materials
opens exciting possibilities to generate nearly one-dimensional
nanostructures \cite{Ugarte,Alvine}. In this context, also the 
filling of silicon dioxide nanochannels \cite{Jarl} and of rolled-up
$InAs/GaAs$ tubes \cite{Deneke} has found great interest. Related 
fluid filling phenomena occur when viscous fluid fronts propagate 
into porous media by spontaneous imbibition \cite{Albert}. On macroscopic
scales, a basic understanding of such capillary rise processes
exists for almost a century \cite{2,3,4,5,6,7}. However, the
applicability of the resulting concepts on the nanoscale has been
the subject of a recent controversy \cite{8,9,10}. In particular,
the conditions under which the Lucas-Washburn equation \cite{2,3}
holds are debated. This equation predicts a $\sqrt{t}$ law for the
rise of the fluid meniscus $H(t)$ in the capillary with time $t$,

\begin{equation}\label{eq1}
H(t) =\left(\frac {\gamma _{LV}R \cos \theta}{2 \eta}\right) ^{1/2}\sqrt{t}.
\end{equation}

Here $\gamma_{LV}$ is the surface tension of the liquid, $\eta$
its shear viscosity, $R$ the pore radius, and $\theta$ the contact
angle between the meniscus and the wall. Eq.(\ref{eq1}) follows by
integration of the differential equation, describing steady state flow,
where the capillary force $2\gamma_{LV}\cos(\theta)/R$ is balanced
by the viscous drag $4\eta d(H(t)/R)^2/dt$ and one assumes that any possible
slip length $\delta \ll R$. Of course,
Eq.~(\ref{eq1}) cannot be true for $t \rightarrow 0$, but can hold
\begin{figure}[thb]
\includegraphics[scale=0.3]{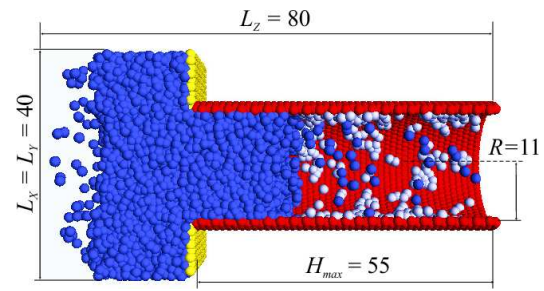}
\caption{Snapshot of fluid imbibition in a capillary at time
$t=1300$ MD time steps after the onset of the process. Fluid atoms
are shown in blue, those of the precursor film - in light blue.
The tube wall is shown in red, and the atoms of the reservoir
adhesive wall - in yellow. For further explanations see text.
\label{snap}}
\end{figure}
only after a (nanoscopically small) transient time (Zhmud et al.
\cite{6} suggest an initial behavior $H(t) \propto t^2$ when the
liquid is accelerated by the capillary forces). However, Mastic et
al. \cite{8} find $H(t)$ rising slower than linear with time, even
for $t \approx 1 ns$, from Molecular Dynamics (MD) simulation of a
simple Lennard-Jones (LJ) fluid. They suggest to slightly correct
Eq.~(\ref{eq1}), replacing $\theta$ by a dynamic contact angle
$\theta(t)$. In contrast, a study of a model for decane molecules
in a carbon nanotube \cite{9,10} yielded a simple linear behavior
$H(t)\propto t$ over a wide range of times, leading to the
conclusion that filling of nanotubes by fluids does not obey the
Lucas-Washburn equation. Experiments so far are inconclusive on
this issue, since the existing work \cite{11,12} deals only with
pores that are at least 1 $\mu m$ wide. Moreover, in narrow nanotubes
an eventual slip at the hydrodynamic boundaries might affect the 
balance of forces by reducing the viscous drag at the tube wall.

The aim of the present letter is to help clarifying the problem
of capillary filling in narrow nanotubes. 
We present simulations of a generic model, varying both
the fluid-wall interaction and the nature of the fluid (simple LJ
particles vs. melt of short polymer chains, respectively). Providing
independent estimates for all the parameters entering
Eq.~(\ref{eq1}), we are able to perform a decisive test of
Eq.~(\ref{eq1}). Since a fluid flowing into a capillary is a
nonequilibrium process, we avoid use of a strictly
``microcanonical protocol'' of our MD simulations, unlike
\cite{9,10}. Using a dissipative particle dynamics (DPD)
thermostat \cite{13}, which does not disturb the hydrodynamic
interactions due to its Galilean invariance, we maintain strict
isothermal conditions, in spite of the heat produced due to the
friction of the flowing fluid. In the real system, the walls of
the nanofluidic device would achieve the thermostating, of course.

\paragraph*{Model description} - The snapshot picture, Fig. \ref{snap}, 
illustrates our simulation geometry. We consider a cylindrical nanotube 
of radius $R=10$, whereby the capillary walls are represented by atoms forming a
triangular lattice with lattice constant $1.0$ in units of the
liquid atom diameter $\sigma$. The wall atoms may fluctuate around
their equilibrium positions at $R+\sigma$, subject to a finitely extensible
non-linear elastic (FENE) potential $U_{FENE}=-15\epsilon_w R_0^2
\ln\left (1- r^2/R_0^2\right ),\;R_0=1.5$. Here $\epsilon_w=1.0
k_BT$, $k_B$ denotes the Boltzmann constant, and $T$ is the
temperature of the system. In addition, the wall atoms interact by
a LJ potential, $U_{LJ}(r)=4\epsilon_{ww} \left[
(\sigma_{ww}/r)^{12}-(\sigma_{ww}/r)^6\right]$, where
$\epsilon_{ww}=1.0$ and $\sigma_{ww}=0.8$. This choice of
interactions guarantees no penetration of liquid particles through
the wall while in the same time the wall atoms mobility
corresponds to the system temperature. In all our studies we use a
capillary length $H_{max}=55$. The right end of
the capillary is closed by a hypothetic impenetrable wall which
prevents liquid atoms escaping from the tube. At its left end the
capillary is attached to a rectangular $40\times 40$ reservoir for
the liquid with periodic boundaries perpendicular to the tube
axis. Although the liquid particles may move freely between the
reservoir and the capillary tube, initially, with the capillary
walls being taken distinctly lyophobic, these particles stay in
the reservoir as a thick liquid film which sticks to the reservoir
lyophilic right wall. The film is in equilibrium with its vapor
both in the tube as well as in the left part of the reservoir. At
a time $t=0$, set to be the onset of capillary filling, we switch
the lyophobic wall-liquid interactions into lyophilic ones and the
fluid enters the tube. Then we perform measurements of the
structural and kinetic properties of the imbibition process at
equal intervals of time. As a simulation algorithm we use the
velocity-Verlet algorithm \cite{14} and DPD thermostat
\cite{13,15} with friction parameter $\xi=0.5$, Heavyside-type
weight functions, and a thermostat cutoff $r_c=2.5\sigma$. The integration
time step $\delta t=0.01t_0$ where $t_0$ is our basic time unit,
$t_0=\sqrt{m\sigma^2/48k_BT}=1/\sqrt{48}$, choosing the particle
mass $m\equiv 1$ and $k_BT\equiv 1$. 

The capillary filling is studied for two basic cases: (i) a simple
fluid interacting via LJ potential with $\epsilon = 1.4$ and
$\sigma = 1.0$, and (ii) a non-Newtonian fluid (a polymer melt)
consisting of short chains of length $N=10$.  The non-bonded
interaction is given by a LJ potential with $\epsilon = 1.4$ and
$\sigma = 1.0$ whereas the bonded forces between chain monomers
result from a combination of FENE and LJ potentials with $\epsilon
= 1.0$ \cite{16}. In both cases the liquid-wall interaction is
given by a LJ potential with strength $\epsilon_{wl}$ which is
varied over a broad range so as to change the wetting
characteristics of the system. All interactions are cut off at
$r_{cut}=2.5\sigma$. By varying the interaction strengths and 
the thermostat parameters, one can change the dynamic properties 
of the test fluids in a wide range. The total number of liquid 
particles is $25000$ while those forming the tube are $3243$.

\begin{figure}
\includegraphics[scale=0.4]{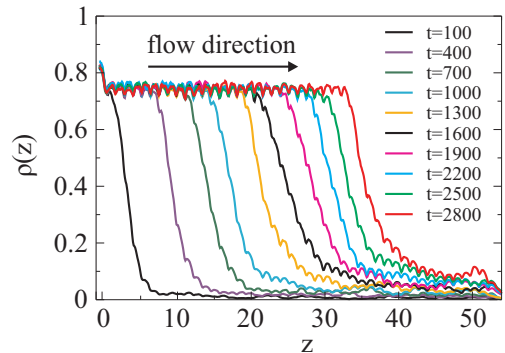}
\caption{Profiles of the average fluid density $\rho(z)$ in the
capillary at various times for the case $\epsilon_{wl}= 1.4,
\epsilon = 1.4$. The small oscillations reflect the corrugated
structure of the wall of the capillary. \label{fig2}}
\end{figure}

\begin{figure}
\includegraphics[scale=0.5]{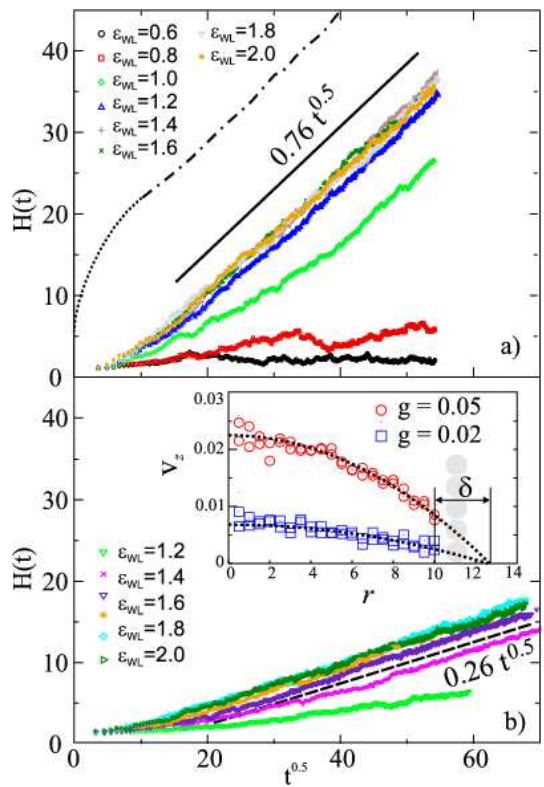}
\caption{Position of the liquid meniscus $H(t)$ for a LJ fluid
(a), and a melt of decanes (b), for various choices of
$\epsilon_{wl}$. The straight lines in (a) - full line, and in (b) - dashed
line, indicate the asymptotic law, Eq.~(\ref{eq1}), in the case of complete wetting.
The topmost dash-dotted curve indicates $H(t)$ for the precursor foot
(calculated from the total number of particles in the first layer
adjacent to the wall of the tube). The initial rise of the
precursor (for $t \leq 100 t_0$) proceeds much faster, however.
The inset in (b) shows the radial velocity variation in a steady flow regime
for two strengths of applied external force $g=0.02$ and $g=0.05$ (see text) 
clearly indicating a slip length $\delta \approx 2.7$.
\label{sqrt}}
\end{figure}

\begin{figure}
\includegraphics[scale=0.4]{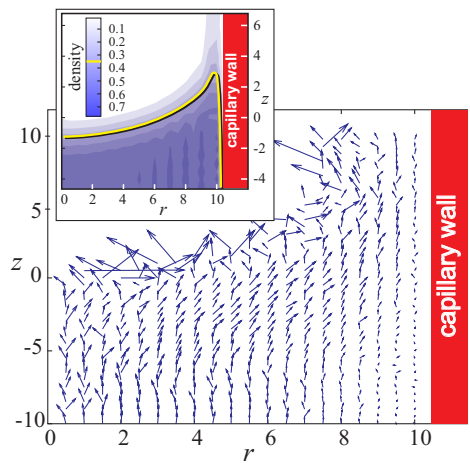}
\caption{Velocity field around the moving meniscus for $\epsilon_{wl}=1.4$. 
Velocities are averaged within a two-dimensional grid, always fixed at $z=0$ to the
actual {\em moving} meniscus position and renormalized according to the current
meniscus speed. The inset shows the LJ-fluid density profile in
the vicinity of the meniscus. The interface position is denoted by a yellow
line. 
\label{fig4}}
\end{figure}

\paragraph*{Simulation results} - Typical data for the time
evolution of the advancing front of the LJ fluid penetrating into
the pore are shown in Fig.~\ref{fig2}. Choosing a constant time
interval ($\Delta t = 300)$ between subsequent profiles in
Fig.~\ref{fig2}, it is already obvious that the interface position
advances into the capillary slower than linear with time! The
profiles $\rho (z)$ at late times become distinctly nonzero far
ahead of the interface position (near the right wall at $z = 55$
where the capillary ends), due to a fluid monolayer attached to
the wall of the capillary: this precursor advances faster then the
fluid meniscus in the pore center, but also with a $\sqrt{t}$ law
(see below).

Fig.~\ref{sqrt}a shows that the time evolution of the meniscus
height $H(t)$ depends very sensitively on the strength of the
wall-fluid interaction (or the contact angle $\theta$,
respectively): for $\epsilon_{WS}=0.6$ and 0.8 only a small number
of fluid particles enter the capillary (since $\theta > 90 $ for
these choices \cite{17}). For $\epsilon_{WS} \geq 1.2$, however,
there is only a short transient up to about $t=250 t_0$ (i.e., a
time still in the nanosecond range), and then a behavior
compatible with Eq.~\ref{eq1} is verified. The initial deviation
from this law seen in Fig.~\ref{sqrt} is not related to the
``dynamic contact angle'' \cite{8}: this would produce a curvature
of opposite sign in Fig.~\ref{sqrt}. However, in the marginal case
$\epsilon_{WL} = 1.0$ a pronounced deviation from Eq.~(\ref{eq1})
occurs: this curve could be (approximately) fitted to a linear
relation, $H (t) \propto t$. It is possible that the results of
Supple and Quirke \cite{9,10} just correspond to such a marginal
case. Finally we emphasize that even the height of the precursor
foot (topmost curve in Fig.~\ref{sqrt}a) advances with a $H(t)
\propto \sqrt{t} $ law.

To test whether the capillary rise behavior of polymers differs
from that of Newtonian fluids, the penetration of a melt of
short flexible polymers (above model (ii)) is shown in
Fig.~\ref{sqrt}b. But apart from a general slowing down
(attributable to the higher viscosity of the polymer melt), the
behavior exhibits the same $\sqrt{t}$-law. 

We have also tested whether the results are affected by possible
simulation artefacts due to insufficient thermostating conditions.
In fact, a slower capillary rise occurs if one uses a ``Langevin
thermostat'' (i.e., an ordinary friction and random noise term
acting on all particles), which violates hydrodynamics \cite{18}.
A similar result applies if the wall atoms are rigid rather than
mobile \cite{17}. But even in these cases the data still follows
the $\sqrt{t}$-law, and also changing details such as
the above parameters chosen for the FENE potential of the wall
atoms does not matter. From the velocity profiles near the moving
meniscus Fig. (\ref{fig4}) it is evident that care is needed for the
temperature equilibration of such non-equilibrium MD simulations
of transient phenomena. The flow is laminar behind the interface,
parallel to the walls of the capillary, with the velocity largest
in the tube center and going to zero close to the walls. Thus our 
simple fluid exhibits evidently stick boundary conditions. However,
in the interface the velocity field bends over into a direction
along the interface, and occasionally particles evaporate into the
gas region. This flow pattern shows that the $H(t) \propto
\sqrt{t}$ must not be confused with a simple diffusive spreading,
of course!

\paragraph*{Comparison with the Lucas-Washburn equation} - For a
test of Eq.~\ref{eq1}, it is crucial to also estimate the
prefactor, of course, to prove that the $\sqrt{t}$ growth is not
just a mere coincidence. 
For the LJ fluid (at density $\rho _\ell = 0.774)$ we
find $\eta \approx 6.34\pm 0.15$, for the polymer melt (at $\rho _\ell =
1.043$) the result is $\eta \approx 205\pm 25$. We derived compatible
values for the viscosity of both fluids also within an equilibrium
Molecular Dynamics simulation by using the correlation function of 
off-diagonal pressure tensor components and the standard Kubo relation \cite{14}. 
From the flat gas-liquid interface observed in the left part of our simulation
box (Fig.~\ref{snap}) we can estimate the surface tension
$\gamma_{\ell v}$ from the anisotropy of the pressure tensor
\cite{19}, $\gamma_{\ell w}= \int dz \{p_{zz}(z)-[p_{xx}(z) +
p_{yy}(z)]/2\}$. This yields $\gamma_{\ell v}=0.735 \pm 0.015$
(LJ) and 1.715$\pm 0.025$ (polymer), respectively.

A consistency check of our results with Eq.(\ref{eq1}) is performed
for the case of complete wetting, $\cos(\theta)=1$, which corresponds
to $\epsilon_{wl} \ge 1.4$ where the data practically collapse on a 
single curve. For the simple LJ-fluid with the tube radius $R=10$
one obtains a slope $H(t)/\sqrt{t} = 0.76\pm 0.02$ which agrees perfectly
with the measured meniscus velocity, cf. Fig. \ref{sqrt}a. For the
polymer melt, in contrast, Eq.(\ref{eq1}) predicts a slope of $0.20$ which 
is considerably less than the observed slopes in Fig. \ref{sqrt}b. 
To clarify this discrepancy we performed MD simulations of steady state
Poiseuille flow of identical melt subject to external force $g$ comparable 
to the capillary driving force. The radial variation of axial velocity
indicates a clear slip-flow behavior, cf. inset in Fig. \ref{sqrt}b, with
a slip length of $\delta \approx 2.7$ which cannot be neglected when compared 
to the tube radius $R=10$. The importance of slip-length in processes in the 
nanoscale range has been emphasized earlier by Barrat and Bocquet \cite{Barrat}.
In the present case the existence of a slip length $\delta$
can be easily accounted for in the Lucas-Washburn result, Eq.(\ref{eq1}),
if one notes that,  according to the definition of a slip length, the drag 
force under slip-flow conditions in a tube of 
radius $R$ and slip length $\delta$ is equal to the viscous drag force for 
a no-slip flow in a tube of effective radius $R+\delta$, that is, to 
$4\eta d(H(t)/(R+\delta))^2/dt$. In both cases the capillary driving
force remains unchanged, $2\gamma\cos(\theta)/R$. Thus one derives a 
modified Lucas-Washburn relationship:
\begin{equation}\label{eq2}
H(t) =\left[\frac {\gamma _{LV}(R+\delta)^2 \cos \theta}{2R \eta} \right]
^{1/2}\sqrt{t}.
\end{equation}
Using Eq.(\ref{eq2}), and the material constants given above, we obtain for 
the slope $H(t)/\sqrt{t} = 0.26\pm 0.02$ which agrees within errorbars with
the observations in Fig. \ref{sqrt}b.
\paragraph*{Conclusions} 
In summary, we have shown that basic concepts of capillarity such as the 
Lucas-Washburn equation, Eq.~\ref{eq1}, work \textit{almost quantitatively} 
even at the nanoscale, both for small molecule fluids and complex fluids such
as short polymer chains. In case of slip-flow, however, we suggest a simple
modification which takes into account the slip length $\delta$. Our new
result, Eq. (\ref{eq2}), restores the consistency of the Lucas-Washburn law
within the framework of the general $\sqrt{t}$ law even in those cases when 
slip-flow cannot be neglected.

\underline{Acknowledgments}: One of us (D.~D.) received
support from the Max Planck Institute of Polymer Research via MPG
fellowship, another (A. M.) received partial support from the
Deutsche Forschungsgemeinschaft (DFG) under project no
436BUL113/130. A.~M. and D.~D. appreciate support by the project "INFLUS",
NMP-031980 of the VI-th FW programme of the EC.


\begin{thebibliography}{99}
\bibitem{17} Full details about our study will be presented
elsewhere later.
\end{thebibliography}
\end{document}